\documentclass[prl,showpacs,twocolumn,amsmath,amssymb,superscriptaddress]{revtex4}

\usepackage{graphicx}
\usepackage{dcolumn}
\usepackage{natbib}

\usepackage{bm}
\usepackage{color}

\newcommand{\Ca}{Ca$_3$Co$_2$O$_6$}
\begin{document}                  
\bibliographystyle{apsrev}

\title{Origin of the long-wavelength magnetic modulation in \Ca}
\author{L. C. Chapon}
 \email{L.C.Chapon@rl.ac.uk}
 \affiliation{ISIS, Rutherford Appleton Laboratory - STFC, OX11 0QX, United Kingdom}

\date{\today}

\begin{abstract}
The origin of the long-wavelength incommensurate magnetic structure of \Ca\ is discussed considering possible inter-chains super-superexchange paths. The experimental value of the propagation vector k=(0,0,$\Delta$) with $\Delta$ $>$ 1 can be reproduced only if one considers the next nearest super-superexchange interaction. A spin-dimer analysis using the Extended-Huckel Tight-Binding method confirms that, despite longer interatomic Co-Co distances, the latter interaction is indeed much stronger. The stability of the observed structure with respect to certain commensurate states is discussed. 
\end{abstract}

\maketitle                        

\begin{figure}[h!]
\includegraphics[scale=0.55,angle=-90]{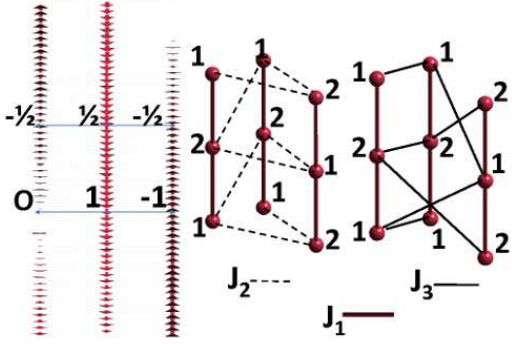}
\caption{(Color online) Left: Sketch of the experimental magnetic structure of \Ca\ showing the longitudinal amplitude modulation for three adjacent chains running along the \textit{c}-axis (25 unit cells are displayed). Specific values of the in-plane magnetic configuration (1,-1,0) and (-$\frac{1}{2}$,$\frac{1}{2}$,-$\frac{1}{2}$) are also shown. Middle and right: Schematic drawing of inter-chain super-superexchange paths between cobalt site 1 (1) and site 2 (2) for J$_2$ (middle) and J$_3$ (right). The ferromagnetic intrachain coupling (J$_1$) is also shown.}
\label{Fig:structure}
\end{figure}

\begin{figure}[h!]
\includegraphics[scale=0.22]{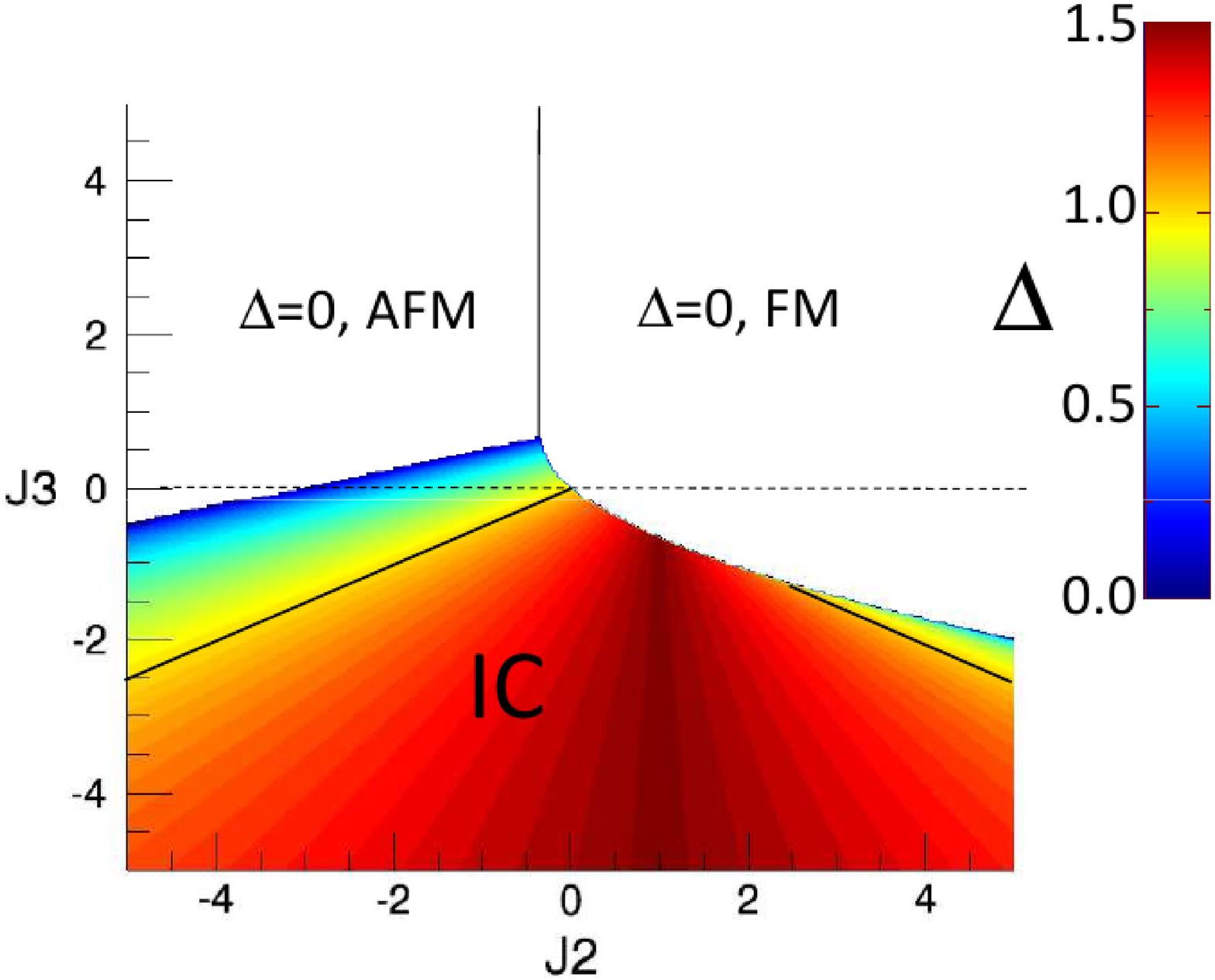}
\includegraphics[scale=0.21,angle=-90]{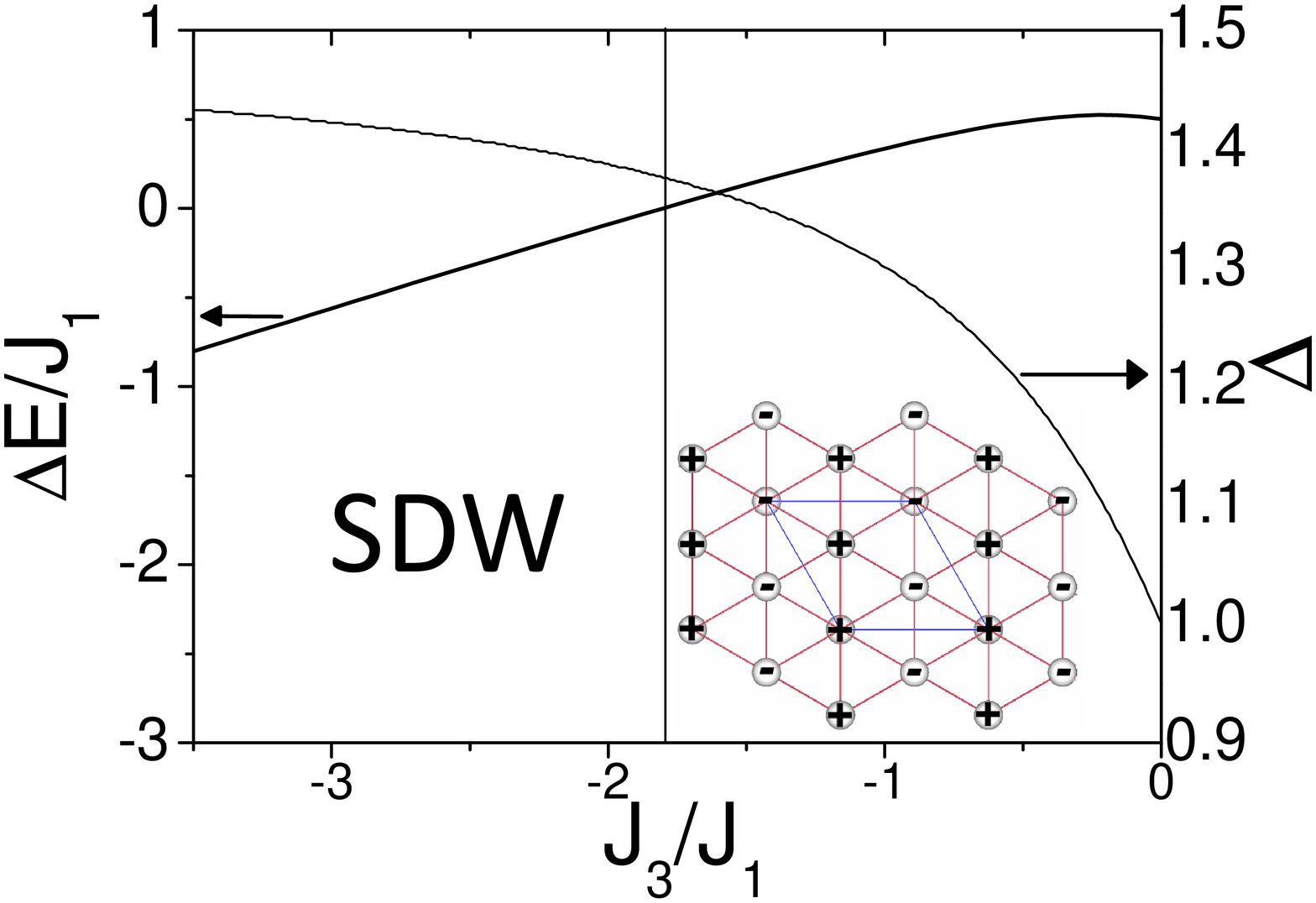}
\caption{(Color Online) \textbf{Top panel:} First ordered state of \Ca\ for isotropic exchange interactions, considering a unitary ferromagnetic intrachain coupling (J$_1$=1) and a wide range of super superexchange interchain coupling parameters (J$_2$, J$_3$). The value of  propagation vector component $\Delta$ is color-coded. \textbf{Bottom panel:} Energy difference between a SDW and the commensurate structure with k=(1/2,1/2,0) shown in the inset as a function of J$_3$/J$_1$ or $\Delta$.}
\label{Fig:exchange}
\end{figure}

The magnetic properties of \Ca\ have attracted considerable interest in the last decade, as this compound is considered as an exact experimental realization of an Ising triangular lattice \cite{PhysRev79357}. Amongst the properties long debated were the nature of the ordered magnetic state and the still ambiguous origin of the steps in the magnetization measurements at low temperature \cite{Kageyama1, Kageyama2, Maignan}, reminiscent of that seen in single-molecule magnets. \\
\indent In the \Ca\ crystal structure only one of the two inequivalent Co$^{3+}$ sites, in trigonal prismatic coordination, is magnetic with a spin-state S=2 and a large orbital contribution. It is well established that these magnetic sites are strongly coupled ferromagnetically within chains running along the \textit{c}-axis and that adjacent chains on the triangular lattice are coupled by weak antiferromagnetic (AFM) interactions in the \textit{ab}-plane \cite{ISI:A1997VY06600011}. However, all studies to date have been undertaken in the framework of a quasi-2D lattice, i.e considering a simple ferromagnetic stacking of AFM triangular planes neglecting the true three-dimensional nature of the exchange interactions. Such approximation, justified in the light of the magnetic structures reported earlier \cite{ISI:A1997VY06600011}, has been challenged by the recent work of Agrestini et al. \cite{ISI:000255457300003,agrestinineutron}. Using first magnetic X-ray diffration \cite{ISI:000255457300003} and then neutron diffraction \cite{agrestinineutron}, these authors have unambiguously established that the magnetic structure is in fact incommensurate, with a modulation along the $c$-axis of very long periodicity ($\sim$ 1000 \AA). Clearly, such a structure can not be stabilized in the quasi-2D limit since the only magnetic interaction along $c$ is ferromagnetic. In order to understand the origin of the proposed ground state it is important to reconsider all possible exchange interactions in the system.\\
\indent In the present communication, I analyze the energy of the magnetic structure considering only isotropic exchange terms and two different super superexchange (SSE) interactions forming helical paths between adjacent chains of the triangular lattice. In this framework, the observed magnetic modulation cannot be explained by considering only the nearest neighbour (noted J$_2$) AFM SSE but requires the presence of a non-vanishing next-nearest AFM SSE terms (J$_3$) which can uniquely stabilize the experimental value of the magnetic propagation vector. A spin-dimer analysis using the Extended-Hukel Tight-Binding (EHTB) method, shows that J$_3$ is indeed the predominant inter-chain coupling term due to a stronger overlap in the molecular orbital involving the Co 3d xz and yz state. Whilst the observed structure corresponds to the first ordered state, for small values of J$_3$ its exchange energy is unfavorable with respect to certain commensurate structures with equal moments, suggesting that the former is stabilized due to a lower entropy.\\     
\indent The magnetic structure reported by Agrestini et al. \cite{agrestinineutron} corresponds to a longitudinal amplitude modulation propagating along the \textit{c}-axis of the hexagonal cell (space group $R\overline{3}m$, hexagonal setting \cite{IUCR}), as illustrated in Fig. \ref{Fig:structure}. In agreement with previous reports, the magnetic moments are aligned along the \textit{c}-axis, fixed by the strong single-ion axial anisotropy \cite{Kageyama2,Maignan}. However, what differs from previous studies is the amplitude modulation or spin-density wave (SDW), i.e the existence of regions in \emph{each chain} with quasi-null ordered moments. It implies that the magnetic configuration within a triangular unit (3 adjacent chains) approaches C1=(+,-,0) \emph{only} at some particular points along the chain (see Fig.\ref{Fig:structure}), unlike the proposed Partially Disordered Antiferromagnet (PDA) structure \cite{metaka,wada} which maintains this configuration along the entire chain. At other lattice points the configuration is almost C2=(+1/2,+1/2,-1/2) (Fig. \ref{Fig:structure}) and, if the structure is truely incommensurate, any other intermediate situations between C1 and C2 are found somewhere in the crystal. Such magnetic arrangement usually originates from competing interactions along the chains in the presence of axial anisotropy since, for isotropic systems, competing interactions would simply lead a non-colinear state with fully ordered moments. The propagation vector in the hexagonal setting is \textbf{k$_H$} = 2$\pi$ (0,0,$\Delta$), with $\Delta$ $\sim$ 1.01 at 18K. $\Delta$ varies with temperature, suggesting a true incommensurability rather than a locking at a particular fractional value. The fact that $\Delta >$ 1 is of particular importance, as will be discussed in the following sections.\\
\indent  For describing the magnetic structure, it is actually more convenient to work in the primitive rhombohedral setting \cite{IUCR}, which will be used in the rest of the communication. In the rhombohedral setting of space group $R\overline{3}m$, the unit-cell dimensions are a=b=c=6.274 \AA\ and $\alpha$=$\beta$=$\gamma$=92.53$^{\circ}$. There are two magnetic Co sites per unit-cell, S$_1$ and S$_2$, at respective fractional coordinates ($\frac{1}{4}$,$\frac{1}{4}$,$\frac{1}{4}$) and ($\frac{3}{4}$,$\frac{3}{4}$,$\frac{3}{4}$). In this setting, the Co-chains are running along the [1,1,1] direction (also the direction of axial anisotropy), and the propagation vector is \textbf{k}=2$\pi$($\frac{\Delta}{3}$,$\frac{\Delta}{3}$,$\frac{\Delta}{3}$). The magnetic structure derived experimentally by Agrestini et al. \cite{agrestinineutron} corresponds to a mode belonging to a single irreducible representation, in agreement with the theory of second-order transition, noted $\Gamma_1$ in Kovalev's notation \cite{Kovalev}. For this mode, the cobalt magnetic moments \textbf{M} on S$_1$ (\textbf{M$_1$}) and S$_2$ (\textbf{M$_2$}), in a unit-cell translated by \textbf{R$_L$}=(R$_x$,R$_y$,R$_z$) with respect ot the zeroth-cell, are written as follows:
\begin{eqnarray}
\label{eq:moment}
\mathbf{M_1(R_L)}&=&\mathbf{M} cos \left( \mathbf{k \cdot R_L} \right) \nonumber \\
\mathbf{M_2(R_L)}&=&-\mathbf{M} cos \left( \mathbf{k \cdot R_L} + \pi\Delta\right)
\end{eqnarray}
where \textbf{M} is a vector pointing along [1,1,1], whose length is the amplitude of the SDW. The energy of this structure can be calculated, in the limit of isotropic exchange interactions, considering the intra-chain ferromagnetic interaction J$_1$, inter-chain super-super exchange (SSE) interactions and a phenomenological single ion anisotropy term DS$^2$. Given the crystallographic parameters, two SSE interactions, mediated through Co-O-O-Co paths, must be taken into account. They correspond to inter-chain nearest (J$_2$) and next-nearest (J$_3$) neighbours, at interatomic Co-Co distances of 5.513 \AA\ and 6.274 \AA\, respectively. Each site (S$_1$ or S$_2$) has six neighbours connected through J$_2$ and six neighbours connected through J$_3$. The list of neighbours is given in Table \ref{table:nn}. These SSE interactions, form helical paths between Co sites of adjacent chains within a triangular motif, as shown in Fig. \ref{Fig:structure}, clearly competing with the ferromagnetic intra-chain exchange, when J$_2$ or/and J$_3$ are antiferromagnetic.   
\begin{table}[h!]
\begin{tabular}{c|c|cc}
SSE Interaction & Site i & Site j & R$_m$ \\
\hline
J$_2$ & S$_1$ & S$_2$ & (-1,0,0) \\
  & S$_1$ & S$_2$ & (0,-1,0) \\
  & S$_1$ & S$_2$ & (0,0,-1) \\
  & S$_1$ & S$_2$ & (-1,-1,0) \\
  & S$_1$ & S$_2$ & (-1,0,-1) \\
  & S$_1$ & S$_2$ & (0,-1,-1) \\
\hline
J$_3$ & S$_{1(2)}$ & S$_{1(2)}$ & (1,0,0) \\
  & S$_{1(2)}$ & S$_{1(2)}$ & (-1,0,0) \\
  & S$_{1(2)}$ & S$_{1(2)}$ & (0,1,0) \\
  & S$_{1(2)}$ & S$_{1(2)}$ & (0,-1,0) \\
  & S$_{1(2)}$ & S$_{1(2)}$ & (0,0,1) \\
  & S$_{1(2)}$ & S$_{1(2)}$ & (0,0,-1) \\
\end{tabular}
\caption{List of super-superexchange paths for \Ca. S$_1$ and S$_2$ refer to the two Co positions in the primitive rhombohedral unit-cell, at fractional coordinate of ($\frac{1}{4}$,$\frac{1}{4}$,$\frac{1}{4}$) and ($\frac{3}{4}$,$\frac{3}{4}$,$\frac{3}{4}$) respectively, while R$_m$ refer to the translation of site j with respect to site i.}
\label{table:nn}
\end{table}
The energy of the magnetic mode found experimentally is easily calculated by summing on all lattice cells (\textbf{R$_L$}). The total energy, taking the convention J$<$ 0 for AFM interactions, is decomposed into a normal (E$_N$) and Umklapp (E$_U$) terms: 
\begin{widetext}
\begin{eqnarray}
\label{eq:energy}
E_N&=&\frac{1}{2}M^2\left(J_1 cos \left(\pi\Delta \right )+3J_2\,cos \left ( \frac{\pi\Delta}{3} \right )-3 J_3 cos\left ( \frac{2\pi\Delta}{3} \right ) + D \right)\\
E_U&=&\frac{M^2}{2N}\sum_{R_l} J_1 \left [ cos \left (2 \mathbf{k \cdot R_L}\right) cos \left( \pi\Delta \right ) \right ]
+ 3J_2 \left [ cos \left (2 \mathbf{k \cdot R_L}\right) cos \left( \frac{\pi\Delta}{3} \right ) \right ] -3 J_3 \left [ cos \left (2 \mathbf{k \cdot R_L}\right) cos\left(\frac{2\pi\Delta}{3} \right ) \right ] \\ \nonumber  
+ && D \left [cos \left (2 \mathbf{k \cdot R_L} \right) \right ]   
\end{eqnarray}
\end{widetext}
When the structure is incommensurate, as found experimentally, one needs to consider only the non-vanishing normal term in equation \ref{eq:energy}. By derivating E$_N$ with respect to $\Delta$, we obtain:
\begin{equation}
J_1 sin \left ( \pi\Delta \right ) + J_2 sin \left ( \frac{\pi\Delta}{3} \right ) -2J_3 sin \left ( \frac{2\pi\Delta}{3} \right ) =0
\label{eq:solution}
\end{equation} 
The parameter $\Delta$ varies in the first Brillouin zone, i.e. $\Delta\le$3/2. Let's consider FM J$_1$ ($>$0) and AFM J$_2$ and J$_3$. By imposing nearest interchain neighbors only (J$_2<$0, J$_3$=0), there are solutions uniquely for $\Delta\le$ 1, since the terms in J$_1$ and J$_2$ must have opposite signs in Eq. \ref{eq:solution}. However, if one considers only FM J$_1$ and J$_3 <$0, it is possible to stabilize solutions with $\Delta > $ 1, as observed experimentally. In particular, reproducing the experimental value of $\Delta\sim$=1.01 requires J$_3$=-0.018J$_1$. \\
Fresard et al. \cite{ISI:000221426200006} already singled-out that J$_3$ is a relevant parameter, based on a shorter O-O distance than that of J$_2$. This argument is often valid, as shown by the work of Whangbo and co-workers in a variety of insulating oxides \cite{ISI:000230259500004}. Based on the crystal structure derived by neutron diffraction at 18K, the O-O distance of 2.908 \AA\ for J$_3$ is indeed shorter than for J$_2$ (2.937 \AA\ ), despite a much longer Co-Co distance (6.274 \AA\ instead of 5.513 \AA). The dihedral angles along the J$_3$ Co-O-O-Co path also maximize the orbital overlap as shown next. The SSE energy can be derived semi-quantitatively by a spin-dimer analysis based on the Extended-Huckel Tight-Binding method using double-$\xi$ slater orbitals for the O s and p states and Co d states \cite{ISI:000230259500004}. Here, each dimer Co$_2$O$_{12}^{18-}$ along J$_2$ and J$_3$ is considered in turn and the exchange energy is directly estimated from the square of the energy difference between the bounding and antibounding levels \cite{ISI:000230259500004}. In the present case, one needs to take into account the multi-electron configuration (d$^4$, S=2 ground state). Since the point symmetry is high ($\overline{3}$m), only the overlap between non-orthogonal orbitals is relevant for the calculations as explained in \cite{ISI:000230259500004}. The spin-dimer analysis has been performed with the program CAESAR 2.0 \cite{CAESAR}, using slater parametrizations for Co and O atoms given in \cite{ISI:000242899400048}. In trigomal prismatic configuration (point group $\overline{3}$m), the Co d orbitals are splitted into one singly degenerate level (z$^2$) and two doubly degenerate levels: (x$^2$-y$^2$,xy) and (xz,yz). The z$^2$ level is fully occupied and does not contribute. The four unpaired electrons occupy the doubly-degenerated levels, the latter lying higher in energy. The spin-dimer analysis shows that the energy difference between bounding and antibounding states involving the (x$^2$-y$^2$,xy) orbitals are not dramatically different for J$_2$ and J$_3$ (considered alone they would lead to J$_2$=1.6J$_3$).
\begin{figure}[h!]
\includegraphics[scale=0.54,angle=-90]{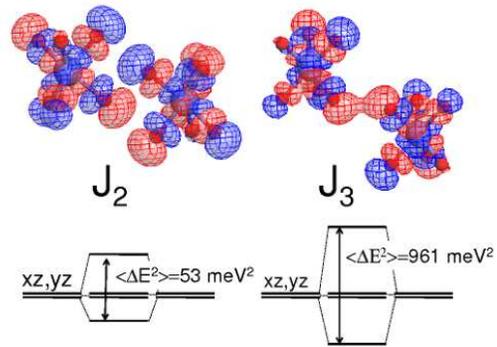}
\caption{Bounding molecular orbitals (involving the Co d$_{xz}$,d$_{yz}$ levels) of the dimer Co$_2$O$_{12}^{18-}$ corresponding to super-superexchange paths along J$_2$ and J$_3$, calculated from the Extended Huckel Tight Binding method. The energy diagram shows the corresponding values of the squared energy difference between bounding and antibounding states (see text for details). }
\label{Fig:spindimer}
\end{figure}
The dimer levels involving the Co (xz,yz) states, form two non bounding molecular orbitals (no interaction within the numerical accuracy) and a pair of bounding and antibounding state, both in the case of the J$_2$-dimer and J$_3$-dimer, as displayed in Fig. \ref{Fig:spindimer}. However the energy gap separating the two latter states is much stronger in the case of J$_3$ (31 meV) than for J$_2$ (7.3 meV), providing a coupling through these orbitals about 20 times stronger for J$_3$. This result, even though semi-quantitative, supports entirely the conclusions conveyed previously in the analysis of the exchange energy, and the initial assumption of Fresard et al. \cite{ISI:000221426200006} highlighting the crucial role of the next nearest SSE interaction in this system. \\
\indent I now turn to the the stability of the SDW structure. One can easily shows that it corresponds to the first ordered state, which can be derived as a function of the propagation vector k=(x,y,z) for various sets of exchange integrals \{$J_{ij}$\}. Here the anisotropy, whose effect is to stabilize a SDW rather than a non-colinear configuration but does not affect the propagation, is not considered. The ground state is obtained by calculating the eigenvector corresponding to the maximum eigenvalues of the Fourier transform of the exchange-integral matrix $\xi_{ij}$ \cite{freiser}:   
\begin{equation}
\xi_{ij}\left( \mathbf{k}, \{ J_{ij} \} \right)= \sum_m J_{ij} \left( \mathbf{R_m} \right) \cdot exp \left(- i \mathbf{k \cdot R_m} \right) 
\end{equation}
The indices i, j refer to the magnetic atoms in a primitive
cell (S$_1$ and S$_2$). J$_{ij}$(Rm) is the isotropic exchange interaction between
the spins of atoms i and j in units cells separated by the
lattice vector R$_m$, as listed in Table \ref{table:nn}. In our case, there are only two magnetic atoms per unit-cell and the two by two Hermitian $\xi_{ij}$ matrix is simply written:  
\begin{equation}
\xi_{ij}\left( \mathbf{k}, \{ J_{ij} \} \right)=\left( \begin{array}{cc}
A & B \\
B$*$ & A \end{array} \right) \\
\end{equation}
where: 
\begin{eqnarray}
A&=&2J_3 \left( cos \left( x \right)+cos\left(y\right)+cos\left(z\right) \right) \nonumber \\
B&=&J_1 \left( 1+e^{i \left( x+y+z \right)} \right) \nonumber \\
&+&J_2 \left( e^{ix}+e^{iy}+e^{iz}+e^{i\left( x+y\right)}+e^{i\left( x+z\right)}+e^{i\left( y+z\right)}\right) \nonumber
\end{eqnarray}
The phase diagram was generated numerically with the program ENERMAG \cite{ISI:000177955700013}, performing a grid-search of \textbf{k} within the first Brillouin zone for various sets of exchange parameters. J$_1$ was fixed to 1, while J$_2$ and J$_3$ were varied between -5 and +5. As expected from the form of $\xi_{ij}$, invariant by permutation of x,y and z, the only type of propagation-vector stabilized varies along a line with x=y=z. The phase diagram is represented in Fig. \ref{Fig:exchange}.
 In two large regions, the propagation vector \textbf{k}=0 is stabilized, with either FM or AFM configurations. In addition, there is a large region of incommensuratbility corresponding to \textbf{k}=2$\pi$($\frac{\Delta}{3}$,$\frac{\Delta}{3}$,$\frac{\Delta}{3}$). As already derived from analysis of the $\Gamma_1$ magnetic mode energy, the line J$_3$=0 stabilizes only $\Delta\le$ 1 and the experimental value of $\Delta=1.01$ requires J$_3\neq$0. However, one can also show that an ordered commensurate (CM) phase with equal moments, not obtained as first ordered state, has a lower exchange energy than the SDW structure. The structure, presented in the inset of Fig. \ref{Fig:exchange} propagates with a single \textbf{k}=(1/2,1/2,0). Every magnetic site is fully ordered and has four out of its six first neighbors aligned antiparallel and two aligned parallel. For values of $\vert J_3/J_1 \vert$ $<$ 1.7, the exchange energy of this structure is lower than the SDW, as presented in Fig. \ref{Fig:exchange}. Moreover, the values of $\Delta$ decreases on cooling according to the resonant X-ray study \cite{ISI:000255457300003}, i.e. the CM structure should be increasingly favourable. This indicate that the SDW structure, as observed, is stabilized only thanks to a smaller configuration entropy. The presence of competing terms in the free energy (exchange and entropy) could explain the crossover regime observed at the so-called freezing temperature T$\sim$18K, temperature below which the coherence length of the SDW is suddenly reduced\cite{agrestinineutron}.\\
\indent In conclusion, I have shown that the long-wavelength magnetic modulation in \Ca\, originates from the existence of interchain AFM super-superexchanges terms forming helical paths running between adjacent chains of the triangular motif. The analysis of the energy and first ordered ground state in the isotropic exchange limit, and a spin-dimer analysis, points to the next-nearest exchange interaction as the critical parameter to reproduce the propagation found experimentally. It is argued that this phase is stabilized thanks to a lower configuration entropy than other structure with constant moments.\\         
\indent I would like to acknowledge Prof. Daniel Khomskii and Prof. Paolo G. Radaelli for frutfull discussions and critical reading of the manuscript.

\end{document}